# Reducing End-to-End Latencies of Multi-Rate Cause-Effect Chains for the LET Model

## Luiz Maia ✉
Rheinland-Pfälzische Technische Universität Kaiserslautern-Landau, Kaiserslautern, Germany
## Gerhard Fohler ✉
Rheinland-Pfälzische Technische Universität Kaiserslautern-Landau, Kaiserslautern, Germany

## — Abstract —

The *Logical Execution Time* (LET) model has been gaining industrial attention because of its timing and data-flow deterministic characteristics, which simplify the computation of end-to-end latencies of multi-rate cause-effect chains at the cost of pessimistic latencies.

In this paper, we propose a novel method to reduce the pessimism in the latencies introduced by LET, while maintaining its determinism. We propose a schedule-aware LET model that shortens the lengths and repositions LET's communication intervals resulting in less pessimistic end-to-end latencies. By adding dependencies between specific task instances, the method can further reduce the pessimism in the latency calculations of the LET model. If needed, e.g., for legacy reasons, our method can be applied to a subset of tasks only.

We evaluate our work based on real world automotive benchmarks and randomly generated synthetic task sets. We compare our results with previous work and the LET model. The experiments show significant reductions of worst-case data age and worst-case reaction latency values.

## 1 Introduction

Modern automotive and avionics systems are highly distributed by design. As a result, devices such as electronic control units (ECUs), monitoring sensors, and actuators have to constantly communicate in order to achieve specific functionalities of the system. Control applications are very sensitive to end-to-end (E2E) communication latencies. For processes present in a controlled application, ensuring execution before a given deadline it is not enough to guarantee the correct behavior of the system. The E2E latencies have to be within a given range, otherwise they can compromise application's control performance significantly [13].

In general, a controlled process consists of a sensor, a controller and an actuator. The sensor samples/reads a given input data and propagates it to the controller, which reads the raw data and processes it. The controller propagates processed data to the actuator, which is responsible for producing the output of the controlled process. This chained sequence of read/write (R/W) operations, called *task chain*, represents a sequence of communication operations between tasks where each task receives data from one or more predecessor tasks. Task chains have been classified as *event chains* when an event produced by a preceding task triggers the execution of its successor; and as *periodic chains* when tasks are periodically activated independently of the execution of their predecessors [21]. A periodic chain is also known as a *cause-effect chain* (CEC). The analysis of whether the E2E latencies of a CEC fulfill required timing constraints is not trivial [9]. The complexity increases when tasks with different periods and multiple data dependencies are present along the CEC. Multi-rate CECs are susceptible to data over/undersampling and, as a consequence, not all instances of the tasks present in a CEC are active contributors in propagating data from one end of the CEC to the other. As a result, data is not easily traceable along the CEC. The complexity of computing E2E latencies of multi-rate CECs increases even further when **23:2      Reducing End-to-End Latencies of Multi-Rate Cause-Effect Chains for the LET Model**considering multi-core systems as tasks can be mapped to different cores and can execute in parallel [20]. On top of that, depending on the execution model considered, e.g., the *Bounded-Execution Time Model (BET)*, the R/W operations do not occur at fixed time points as they depend on when tasks begin and end execution [8]. Thus, the R/W operations can take place at different time points for different instances of the same task, which results in non-deterministic accesses to shared variables used for inter-task communication.

The Logical Execution Time (LET) model [10] was adopted as a way to overcome the complexity of computing E2E latencies of multi-rate CECs. By defining specific time points for tasks' R/W operations, the LET model enables: deterministic accesses to shared variables, deterministic timing and data-flow, as well as composability [8]. In LET, the R/W operations can only occur at the boundaries of the so-called *communication interval* ($L$) [16], which is considered to be equal to tasks' period. As a result, LET is a schedule-agnostic model, since the schedule of the tasks does not affect when the R/W operations occur. However, limiting the R/W operations to happen only at the boundaries of $L$ results in pessimistic E2E latencies. Becker et. al [2], Kordon and Tang [11] and Martinez et. al [15], focused on proposing methods to compute worst/best-case E2E latencies of CECs applying the LET model. Ernst et. al [5] focused on extending the LET model to the system level on distributed systems. In [14], Martinez et. al proposed to use an offset-aware LET model to reduce E2E latencies. In [4], Bradatsch et al. proposed a worst-case response time LET model.

Motivated by the WATERS industrial challenge proposed in [9] and based on the Real World Automotive Benchmarks presented by Kramer et al. in [12], we address the problem of pessimistic E2E latencies present in multi-rate CECs applying the LET model. We propose a method to be used during later design phases when more concrete knowledge about the underlying system is available. Given a schedulable task set in a multi-core system, our method sets schedule-aware intervals that change the lengths and positions of the communication intervals and reduce the pessimism of the E2E latencies. On top of that, we add dependencies between specific task instances in order to further reduce the E2E latencies. Without losing the properties of the LET model, we propose a method that is legacy-aware and can be applied individually to selected tasks and/or CECs during later design phases.

The remainder of the paper is structured as follows. In Section 2 we present related work, while in Section 3 we briefly present our application model and the required assumptions. In Section 4 we motivate our work and introduce the base LET model. Section 5 presents our method to reduce the E2E latencies of multi-rate CECs. In Section 6 we evaluate our work using the Real World Automotive Benchmarks [12] and the paper is concluded in Section 7.

## 2   Related Work

Feiertag et al. [6] identified four E2E timing semantics that characterize different timing latencies. Of these four, the two most commonly considered latencies when analyzing a multi-rate CEC are: *reaction latency* and *data age*. The reaction latency measures the *reactivity* of the system to new inputs, it is the maximum delay between a valid input until the *first* output based on such input is produced by the last task in the CEC. Data age measures the *freshness* of data, it is the maximum delay between a valid input until the *last* output based on such input is produced by the last task in the CEC. Becker et al. proposed in [1] to use job-level dependency (JLD) as a way to control data propagation and E2E latencies in multi-rate CECs applying the BET model. By setting precedence constraints between specific jobs, Becker et al. eliminate data propagation paths which exceed the timing requirements of the multi-rate CECs. A method to compute the worst/best-case data age is also presented in [1].

The LET model was first introduced in [10] as part of the Giotto programming language in the context of time-triggered tasks. Biondi et al. presented in [3] an implementation of the LET model on actual multi-core platforms for automotive systems. Biondi et al. also proposed equations to identify which instances of producer/consumer tasks contribute to data propagation in multi-rate CECs applying the LET model. In [17], Pazzaglia et al. used LET to enforce causality and determinism as a way to control accesses to shared memory and optimize the functional deployment on multi-core platforms. In [2], Becker et al. extended their previous work in [1] to also consider LET. However, in [2], JLDs are not applied to the LET model and the communication interval is assumed to be equal to a task's period.

In [15], Martinez et. al improved the work proposed in [2] by providing tighter bounds on the computation of the E2E latencies of multi-rate CECs. Kordon and Tang proposed in [11] a mathematical algorithm to determine the worst-case data age of multi-rate CECs. In [14], Martinez et al. presented an offset-aware LET analysis to improve the real-time performance of an automotive engine control system. Bradatsch et al. proposed in [4] a method to reduce the data age of multi-rate CECs applying the LET model by setting the communication intervals equal to the worst-case response time (WCRT). In [5], Ernst et al. extended the LET model to the system level and enabled LET to be used in distributed systems. Ernst et al. abstract from the non-zero communication time between the different end systems by defining intermediate tasks, which are LET tasks with the communication intervals equal to the worst-case transmission time (WCTT).

Differently from [2] and [1], we do not limit the use of JLDs to the BET model. By setting schedule-aware intervals and allowing JLDs in the LET model, we obtain less pessimistic E2E latencies and improve the results shown in [4].

## 3 Application Model and Assumptions

We consider a multi-core system composed of identical cores and a task set $\Gamma$, which consists of $n$ periodic and independent real-time tasks, $\Gamma = \{\tau_1, ..., \tau_n\}$. Each $\tau_i \in \Gamma$, $0 < i \leq n$, is a tuple $(C_i, T_i)$, where $C_i$ represents the worst-case execution time (WCET) and $T_i$ its period. We assume that tasks have implicit deadlines, i.e. $D_i = T_i$, and do not overrun. All tasks in $\Gamma$ are released simultaneously at $t = 0$. A task instance (job) $\tau_{i,k}$ represents the $k^{th}$ instance of task $\tau_i$ within the hyperperiod ($HP$), where $0 \leq k < \frac{HP}{T_i}$, and $HP = lcm_{\forall \tau_i \in \Gamma}\{T_i\}$. Tasks are mapped offline to specific cores and migration is not allowed.

We assume that all tasks in $\Gamma$ follow the LET model and that a task can only start execution after its initial read operation in the beginning of the communication interval ($L$) has taken place. Communication between tasks with the same or different periods happens through shared variables, where only the last written data is stored. Inter-task communication between tasks mapped to the same core happens via the use of shared variables located on a local memory, while tasks mapped to different cores use shared variables located on a shared memory. We assume that the shared variables are initialized with default values if no data was written to them. If the data written in a shared variable by $\tau_i$ is read by $\tau_j$, $j \neq i$, we consider that $\tau_j$ consumes the data produced by $\tau_i$. The $\rightarrow$ operator means that $\tau_j$ acts as a consumer/reader task, while $\tau_i$ as a producer/writer task, i.e. $\tau_i \rightarrow \tau_j$. If $\tau_i$ is accessing a shared variable to perform a read operation while $\tau_j$ is accessing the same variable to perform a write operation, we consider that the write operation always precedes the read operation. As in [14], for a given shared variable, we allow multiple readers and a single writer.

We model the communication between tasks as directed acyclic graphs. A directed acyclic graph (DAG) $G = (N,A)$, consists of a set of nodes $N$ and a set of arcs $A$. A node $N_i$, $0 < i \leq n$, represents a task $\tau_i$. An arc $A_{i,j}$, represents the inter-task communication between $\tau_i$ and $\tau_j$. For $G = (N,A)$, there is a finite set of paths between source and sink nodes. We analyze each path as a linear, acyclic CEC that does not consist of junctions [1].

We define a CEC as $E$, $E$ being an ordered sequence of tasks $(\tau_i^1 \to ... \to \tau_j^\eta)$, where $\eta$ is the number of tasks constituting $E$. Task $\tau_i^1$ represents the first (writing) task in the CEC $E$ and $\tau_j^\eta$ the last (reading) task in $E$. The hyperperiod of the tasks present in the CEC $E$ is $HP_E = lcm_{\forall \tau_i \in E}\{T_i\}$. We assume in this work that incoming inputs to $E$ are periodic, and that every new input to $E$ always arrives synchronously with the read operation of all $\tau_{i,k}^1 \in \tau_i^1$.

# 4 The Logical Execution Time Model (LET) and Motivating the Proposed Method

Timing and data-flow determinism are the main benefits of the LET model [8] and are possible due to the decoupling of the *communication phase* from the *execution phase*. The execution phase represents the time interval between the start of task's execution until its completion. The communication phase consists of the points in time at which a task $\tau_i$ accesses a shared variable either for reading or writing. The read/write (R/W) operations of a task $\tau_i$ can only occur at the boundaries of the communication interval ($L_i$). Since $L_i = T_i$, the boundaries of $L_i$ correspond to the release times of $\tau_{i,k}$, e.g., begin at $kT_i$ and end at $(k+1)T_i$. By fixing the R/W operations to only occur at specific points in time (w.r.t. $T_i$), $L_i$ can be seen as an abstraction layer that makes the R/W operations independent from the actual execution of $\tau_i$. No matter when or for how long $\tau_i$ executes (execution phase), the R/W operations (communication phase) are always performed at the same specific time points (w.r.t. $T_i$).

At the beginning of $L_i$, $\tau_i$ performs its read operation and copies from a shared variable to a local variable all the data necessary for its execution. As a result, even if the content of the shared variable accessed by $\tau_i$ has been overwritten during $\tau_i$'s execution, data consistency is maintained with respect to the data accessed by $\tau_i$. At the end of $L_i$, $\tau_i$ performs its write operation and copies the data stored in the local variable to the shared variable that will be accessed by the next task in the CEC. LET assumes that the R/W operations are atomic and are performed in zero time, which means that all written values are instantaneously available to all read operations [8].

As presented in [14], for a given inter-task communication, $\tau_i \to \tau_j$, between two tasks applying the LET model, a *publishing point* ($P_{W,R}^n$) represents the point in time where the content of the variable shared between the writer task $\tau_i$ and the reader task $\tau_j$ is updated, $W$ being the index of $\tau_i$, $R$ the index of $\tau_j$ and $n \geq 0$ [14]. After $P_{W,R}^n$, no other write operations of $\tau_i$ will occur before the next read operation of $\tau_j$. A *publishing instance* represents the instance of $\tau_i$ that updates the variable shared between $\tau_i$ and $\tau_j$. A *reading point* ($Q_{W,R}^n$) represents the point in time where an instance of the reader task $\tau_j$, known as *reading instance*, reads for the first time the new data published by the preceding $P_{W,R}^n$ point. Depending on the periods of writer and reader tasks, not all time points where $\tau_i$ performs its write operations result in publishing points and not all time points where $\tau_j$ performs its read operations result in reading points [14]. Equations 1 and 2 show how to compute the publishing and reading points of an inter-task communication, $\tau_i \to \tau_j$, between two tasks applying the LET model. The proof for 1 and 2 is presented in [14].

$$P^n_{W,R} = \left\lfloor \frac{n * max(T_W, T_R)}{T_W} \right\rfloor T_W \quad (1) \quad Q^n_{W,R} = \left\lceil \frac{n * max(T_W, T_R)}{T_R} \right\rceil T_R \quad (2)$$

Consider the task set $\Gamma = \{\tau_1, \tau_2, \tau_3\}$ and the CEC $E = (\tau_1 \rightarrow \tau_2 \rightarrow \tau_3)$, where $\tau_1 :< 1, 3 >, \tau_2 :< 1, 5 >$ and $\tau_3 :< 1, 3 >$. Tasks follow the LET model and run on the same core in order to facilitate the explanation of this example. However, this example can be easily applied to a multi-core system. Assume that new inputs to $E$ are periodic and always arrive synchronously with the read operations of $\tau_1$ at the beginning of $L_1$. Figure 1 shows how data propagates through $E$ within the first $HP_E$. At the beginning of every communication interval, $L_{\tau_1} = T_1$, job $\tau_{1,k}$ copies a new input value to its local variable. At the end of every $L_1$, $\tau_{1,k}$ copies its execution results from its local variable to the variable shared between $\tau_1$ and $\tau_2$. Data propagates through $E$ according to the publishing and reading points of each task in $E$. For $\tau_1 \rightarrow \tau_2$, the data read at $t = 0$ propagates from $\tau_1$ to $\tau_2$ at the publishing point $P^1_{\tau_1,\tau_2} = 3$, but it is only read by $\tau_2$ at the reading point $Q^1_{\tau_1,\tau_2} = 5$ because $L_{\tau_2} = T_2$. Data continues to propagate through $E$ from $\tau_2$ to $\tau_3$ at $P^2_{\tau_2,\tau_3} = 10$, which is read by $\tau_3$ at $Q^2_{\tau_2,\tau_3} = 12$. The first output of $E$ happens at $t = 15$ when $\tau_3$ writes its execution results. In this case, both reaction latency and data age for the input value read at $t = 0$ are equal to 15.

Figure 1 shows that, although task $\tau_1$ released five jobs, only one value propagated until the end of $E$ during one $HP_E$. The input values read at $t = \{3, 9\}$ did not propagate through $E$ and were never consumed by the next tasks in the chain. Their values were overwritten by the inputs read at $t = \{6, 12\}$ respectively. Note that the input read at $t = 6$ will continue its propagation through $E$ during the next $HP_E$. The next output will be produced at $t = 18$, which results in a reaction latency of 12 and a data age of 15 for the input value read at $t = 6$. Table 1 shows the E2E latencies for the first six outputs that propagated through the entire CEC $E$ based on the inputs read by $\tau_1$ starting from $t = 0$ until $t = 15$. Note that the outputs produced at $t = \{3, 6, 9, 12\}$, which are based on values that did not propagate through the entire CEC are not shown in Table 1 because they do not represent complete E2E latencies during the first $HP_E$.

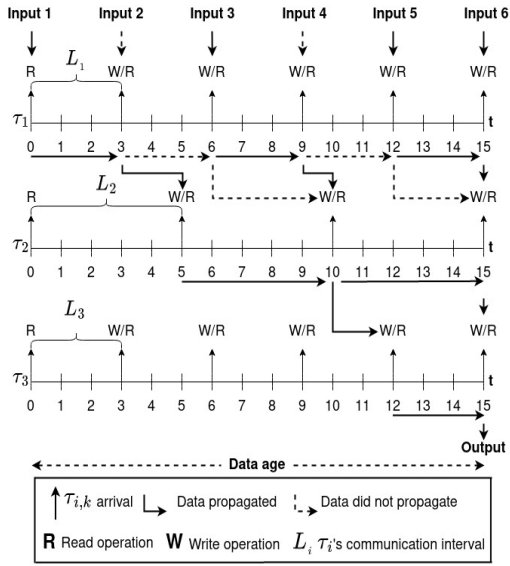

**Figure 1** Data propagation in $E$ with $L_i = T_i$

| Input at t: | Output at t: | E2E latency: |
|---|---|---|
| 0 | 15 | 15 |
| 6 | 18 | 12 |
| 6 | 21 | 15 |
| 12 | 24 | 12 |
| 12 | 27 | 15 |
| 15 | 30 | 15 |

**Table 1** E2E latencies of $E$ with $L_i = T_i$

The E2E latencies shown in Table 1 can be improved by setting tighter bounds for both ends of the communication intervals, while still guaranteeing timing and data-flow determinism. At runtime, even if task $\tau_i$ executes early during its period $T_i$, the result of $\tau_i$'s execution is only propagated through the CEC at the end of $L_i$. Considering $L_i = T_i$ is a conservative approach that makes LET schedule-agnostic, which can be a benefit during early design phases when there are no decisions about the system platform or the scheduling algorithm. Later during design, when more concrete knowledge about the underlying system and scheduling algorithm is available, there is the possibility that E2E latencies can be improved without losing the deterministic properties of LET. In Section 5, we introduce our method to make communication intervals schedule-aware and have less pessimistic E2E latencies, while keeping the deterministic properties of LET.

## 5     Manipulating LET Communication Intervals to Reduce End-to-End Latencies

As discussed in Section 4, multi-rate CECs applying the LET model have timing and data-flow determinism, but pessimistic E2E latencies. In this section, we propose a method that makes the communication intervals schedule-aware. Instead of setting $L_i = T_i$, $\forall \tau_i \in \Gamma$, we set new points in time for the R/W operations to occur, which implies changing the length and position of $L_i$. However, we maintain the deterministic characteristics of the LET model. Our method draws inspiration from the concept of *target windows* [7, 18] which were initially used in the context of increasing the acceptance rate of aperiodic tasks in time-triggered systems. In [18], a target window is defined as the time interval in which a task can execute at any point in time without compromising its timing and precedence constraints. This definition resembles the concept of a communication interval, $L_i$, where the timing constraints of $\tau_i$ are respected as long as $\tau_i$ executes within $L_i$.

### 5.1    Defining Schedule-Aware Intervals

In order to make $L_i$ schedule-aware, $\forall \tau_i \in \Gamma$, we set $L_i$'s length and position equal to the interval $I_i$, where the length of $I_i$ is $C_i \leq |I_i| \leq T_i$. The time points $begin(I_i)$ and $end(I_i)$ are the points in time (w.r.t. $T_i$), where the R/W operations take place respectively. Thus, $begin(I_i)$ and $end(I_i)$ delimit the boundaries of $I_i$. The length and position of $I_i$ is calculated based on the notion of *execution intervals* $EI_{i,k}$, $\forall \tau_{i,k} \in \tau_i$. We define an execution interval $EI_{i,k}$ of job $\tau_{i,k}$, as the time interval between the *earliest reading point* ($erp_{i,k}$) and the *last writing point* ($lwp_{i,k}$). The earliest reading point $erp_{i,k}$ represents the point in time (w.r.t. $T_i$), when $\tau_{i,k}$ is ready to perform its read operation and start its execution. The last writing point $lwp_{i,k}$ represents the point in time (w.r.t. $T_i$), when $\tau_{i,k}$ completes its execution and writes its results. Each $\tau_{i,k}$ can have different values for $erp_{i,k}$ and $lwp_{i,k}$ depending on when each $\tau_{i,k}$ executes between its release and deadline. At runtime, not all jobs of $\tau_i$ necessarily have the same execution interval. Some jobs may execute early during their period, while others may execute later. Thus, it is not possible to set $L_i = EI_{i,k}$, because this results in non-deterministic points in time for the R/W operations to occur. In order to keep the determinism of LET when setting $L_i = I_i$, it is necessary to ensure that all $\tau_{i,k} \in \tau_i$ have a common time-interval for performing their R/W operations.

By analyzing a schedulable task set according to a given scheduling policy, we can identify, $\forall \tau_{i,k} \in \tau_i$, the minimum $erp_{i,k}$ and the maximum $lwp_{i,k}$. Using Equations 3 and 4, we can compute, $\forall \tau_i \in \Gamma$, $begin(I_i)$ and $end(I_i)$. Equations 3 and 4 define the common time-interval $I_i$ (w.r.t. $T_i$), where $\forall \tau_{i,k} \in \tau_i$ can perform their R/W operations without compromising the determinism of the LET model. By setting $L_i = I_i$ and $begin(I_i)$ equal to the minimum $erp_{i,k}$, we ensure that independently of when $\tau_{i,k}$ starts its execution, early or late (w.r.t. $T_i$), there is no $\tau_{i,k} \in \tau_i$ that performs its read operation before time point $begin(I_i)$. By computing the maximum $lwp_{i,k}$, we ensure that even if a $\tau_{i,k}$ finishes its execution early or late (w.r.t. $T_i$), there is no $\tau_{i,k} \in \tau_i$ that performs its write operation after time point $end(I_i)$.

$$begin(I_i) = \min_{\forall \tau_{i,k} \in HP} erp_{i,k} \quad (3) \quad end(I_i) = \max_{\forall \tau_{i,k} \in HP} lwp_{i,k} \quad (4)$$

Let us consider the same task set $\Gamma$ and CEC $E$ presented in Section 4. We compute $begin(I_i)$ and $end(I_i)$, $\forall \tau_i \in \Gamma$, using Equations 3, 4 and the scheduling points shown in Figure 2. Table 2 shows the computed values for $begin(I_i)$ and $end(I_i)$ (w.r.t. $T_i$). By setting $L_i = I_i$, we shorten the length of $L_i$ and reposition it within the time interval $T_i$. As a result, the read operations are postponed and the write operations are preponed (w.r.t. $T_i$). Thus, producer and consumer tasks along the CEC are able to read (resp. write) fresher data values than with $L_i = T_i$.

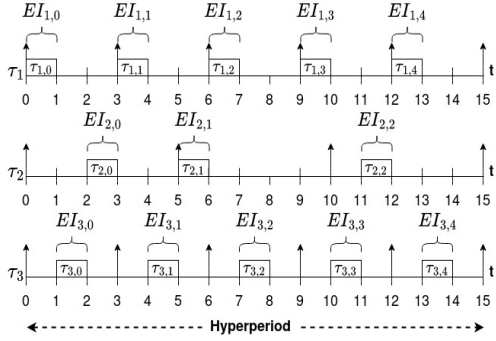

**Figure 2** Schedule for $\Gamma = \{\tau_1, \tau_2, \tau_3\}$ according to EDF

| Schedule-Aware Interval $I_i$ | $begin(I_i)$ | $end(I_i)$ |
|---|---|---|
| $I_1$ | 0 | 1 |
| $I_2$ | 0 | 3 |
| $I_3$ | 1 | 2 |

**Table 2** $begin(I_i)$ and $end(I_i)$ values for tasks $\tau_1$, $\tau_2$ and $\tau_3$

As we reduce and reallocate communication intervals by considering schedule-aware intervals, the assumption that $L_i = T_i, \forall \tau_i \in \Gamma$, in Equations 1 and 2 does not hold anymore. Inspired by the work in [14], we define in Theorems 1 and 2 new equations to compute $P_{W,R}^n$ and $Q_{W,R}^n$ points of multi-rate CECs applying the LET model with schedule-aware intervals. Although extensions to consider task offset are shown in [14], they are only applicable with $L_i = T_i$. Note that in Theorems 1 and 2, when $n = 0$, points $P_{W,R}^0$ and $Q_{W,R}^0$ are not necessarily at $t = 0$ as with $L_i = T_i$. For $L_i = I_i$, it is possible that the points $P_{W,R}^0$ and $Q_{W,R}^0$ are negative numbers for $n = 0$, which means that they are points before the release of $T_W$ (resp. $T_R$) and are not to be considered. However, disregarding the negative points for $n = 0$ does not affect the calculation of the next $P_{W,R}^n$ and $Q_{W,R}^n$ points, since the negative points roll around during the $HP_E$. We prove Theorems 1 and 2 similarly as in [14].

▶ **Theorem 1.** *Given an inter-task communication ($\tau_i \to \tau_j$) between two tasks applying the LET model with schedule-aware intervals, where $T_W \leq T_R$, W being the index of $\tau_i$ and R the index of $\tau_j$, the publishing and reading points can be computed as*

$$P_{W,R}^n = \lfloor \frac{nT_R + begin(I_R) + (T_W - end(I_W))}{T_W} \rfloor T_W - (T_W - end(I_W))$$
$$Q_{W,R}^n = nT_R + begin(I_R)$$

**Proof.** If the period of the writer task is smaller or equal to the period of the reader task, e.g., $T_W \leq T_R$, then there is one publishing and one reading point for each reading instance of the reader task. Reading points correspond to each release of the reader tasks $(nT_R)$ plus the beginning of the interval $I_R$, $begin(I_R)$. The publishing points correspond to the last write operation of the writer task before the aforementioned reading point, i.e. $P_{W,R}^n = \lfloor (nT_R + begin(I_R) + (T_W - end(I_W)))/T_W \rfloor T_W - (T_W - end(I_W))$. ◀

▶ **Theorem 2.** *Given an inter-task communication $(\tau_i \rightarrow \tau_j)$ between two tasks applying the LET model with schedule-aware intervals, where $T_W \geq T_R$, W being the index of $\tau_i$ and R the index of $\tau_j$, the publishing and reading points can be computed as*

$$P_{W,R}^n = nT_W - (T_W - end(I_W))$$
$$Q_{W,R}^n = \lceil \frac{nT_W - (T_W - end(I_W)) - begin(I_R)}{T_R} \rceil T_R + begin(I_R)$$

**Proof.** If the period of the writer task is larger or equal to the period of the reader task, e.g., $T_W \geq T_R$, then there is one publishing and one reading point for each publishing instance of the writer task. Publishing points correspond to the previous write operation of the writer task before each of its releases. The reading points correspond to the next read operation of the reader task after the aforementioned publishing point, i.e. $Q_{W,R}^n = \lceil (nT_W - (T_W - end(I_W)) - begin(I_R))/T_R \rceil T_R + begin(I_R)$. ◀

Note that if we consider $begin(I_i) = 0$ and $end(I_i) = T_i$ as when $L_i = T_i$, the equations in Theorems 1 and 2 go back to their original form as presented in [14]. Hereafter, we propose new equations for computing the worst-case data age and worst-case reaction latency. Given a CEC $E$ composed of an ordered sequence of $\eta$ tasks $(\tau_i^1 \rightarrow ... \rightarrow \tau_j^\eta)$ applying the LET model with schedule-aware intervals, there is a finite number of communication paths in a hyperperiod. As in [15], we call these communication paths as *basic paths*. The publishing and reading points between two tasks in the $n^{th}$ basic path of $E$ are $\dot{P}_{W,R}^n$ and $\dot{Q}_{W,R}^n$ respectively. Note that $\dot{P}_{W,R}^n$ and $\dot{Q}_{W,R}^n$ are not necessarily equal to $P_{W,R}^n$ and $Q_{W,R}^n$ [15]. The boundaries of the $n^{th}$ basic path of $E$ are $\dot{P}_{W^1,R^2}^n$ and $\dot{Q}_{W^{\eta-1},R^\eta}^n$, where the superscript $\{1, ..., \eta\}$ represents the position of the task in $E$, e.g., $W^1$ represents the index of the first task $\tau_i$ in $E$ and $R^\eta$ the index of the last task $\tau_j$ in $E$. By computing $\dot{P}_{W^{\eta-1},R^\eta}^n$ for every $\dot{Q}_{W^{\eta-1},R^\eta}^n$ within $HP_E$ and the corresponding preceding $\dot{Q}_{W,R}^n$ (resp. $\dot{P}_{W,R}^n$) points for all the remaining pairs of communicating tasks comprising $E$, the value of the $\dot{P}_{W^1,R^2}^n$ can be obtained [15]. Note that paths starting from the same publishing point $\dot{P}_{W^1,R^2}^n$ as a previous computed path are not be considered [15]. The length of the $n^{th}$ basic path of $E$ can be computed as: $\theta_E^n = \dot{Q}_{W^{\eta-1},R^\eta}^n - \dot{P}_{W^1,R^2}^n$. As the read operation of $\tau_i^1$ triggers the activation of $E$, the data age $\alpha$ related to the $n^{th}$ basic path of $E$ can be computed using Equation 5.

$$\alpha^n = (end(I_i^1) - begin(I_i^1)) + \theta_E^n + \varphi_E^n - begin(I_j^\eta) - (T_j^\eta - end(I_j^\eta)) \quad (5)$$

In Equation 5, term $(end(I_i^1) - begin(I_i^1))$ represents the interval between the reading and write operations of $\tau_i^1$ in $E$. Term $\theta_E^n$ represents the length of the $n^{th}$ basic path. The interval $\varphi_E^n = (\dot{Q}_{W^{\eta-1},R^\eta}^{n+1} - \dot{Q}_{W^{\eta-1},R^\eta}^n)$ represents the interval between the end of $n^{th}$ basic path and the end of the next $n+1^{th}$ basic path. Since we consider $L_j^\eta = I_j^\eta$, $begin(I_j^\eta)$ and $end(I_j^\eta)$ are not necessarily equal to 0 and $T_j^\eta$ respectively, as with $L_j^\eta = T_j^\eta$. Thus, we have to subtract the difference $begin(I_j^\eta)$ and $(T_j^\eta - end(I_j^\eta))$ from interval $(\dot{Q}_{W^{\eta-1},R^\eta}^{n+1} - \dot{Q}_{W^{\eta-1},R^\eta}^n)$ in order to correctly represent the interval between $\dot{Q}_{W^{\eta-1},R^\eta}^n$ and the last output of $\tau_j^\eta$ based on $\dot{P}^n_{W^1,R^2}$. The worst-case data age of the $E$ is the maximum data age value among all the basic paths within $HP_E$, $\alpha(E) = \max\limits_{\forall n \in HP_E} \alpha^n$.

Equation 6 shows how to compute the reaction latency considering that inputs to $E$ are periodic and always arrive synchronously with the read operation of the first task $\tau_i^1$ in $E$. The worst-case reaction latency is the maximum reaction latency value among all the basic paths within $HP_E$, $\delta(E) = \max\limits_{\forall n \in HP_E} \delta^n$.

$$\delta^n = (end(I_i^1) - begin(I_i^1)) + \theta_E^n + (end(I_j^\eta) - begin(I_j^\eta)) \tag{6}$$

Figure 3 shows how schedule-aware intervals affect the E2E latencies and data propagation through the CEC $E$ considered in Section 4. As demonstrated in Figure 3, by setting $L_i = I_i$, $\forall \tau_i \in \Gamma$, within one $HP_E$, two inputs read by $\tau_1$ starting from $t \geq 0$ managed to propagate throughout the entire CEC $E$ instead of only one as shown in Figure 1, with $L_i = T_i$. Note that in Figure 3 the end of the timeline is at end of the first $HP_E$, $t = 15$, for this reason we do not show the data age for the input value read at $t = 9$, because another output based on this input is produced at $t = 17$. The worst-case data age occurs for the input read at $t = 12$, which affects the output of $E$ from $t = 20$ until $t = 23$. Table 3 shows the E2E latencies for the first six outputs that propagated through the entire CEC $E$ based on the inputs read by $\tau_1$ within the time interval of two $HP_E$. Note that the outputs produced at $t = \{2, 5, 8\}$, which are based on values that did not propagate through the entire CEC are not shown in Table 3 because they do not represent complete E2E latencies.

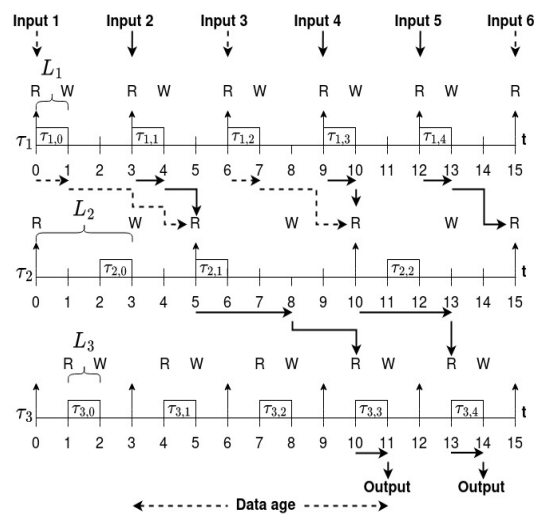

| Input at t: | Output at t: | E2E latency: |
|---|---|---|
| 3 | 11 | 8 |
| 9 | 14 | 5 |
| 9 | 17 | 8 |
| 12 | 20 | 8 |
| 12 | 23 | 11 |
| 18 | 26 | 8 |

**Table 3** E2E latencies of $E$ with $L_i = I_i$

**Figure 3** Data propagation in $E$ with $L_i = I_i$

Figure 3 and Table 3 show the improvements of setting $L_i = I_i$ instead of $L_i = T_i$. One improvement is that now fresher input values produced by task $\tau_1$ at the beginning of $E$ are able to propagate until the end of the chain. Another improvement is that E2E latencies are reduced as result of postponed read operations and preponed write operations. The reduction of the E2E latencies can be observed by comparing Tables 1 and 3. On top of that, the worst-case reaction latency is reduced from 15 to 8 and the worst-case data age from 15 to 11. In Section 6 we apply our method to different configurations of multi-rate CECs in order to evaluate our method of using schedule-aware intervals.

## 5.2 Manipulating Schedule-Aware Intervals by Adding Job-Level Dependencies

In Section 5.1, we showed that by using schedule-aware intervals, fresher data values propagate throughout the CEC and result in less pessimistic E2E latencies compared with $L_i = T_i$. In order to further improve the E2E latencies, we change the position and length of $I_i, \forall \tau_i \in \Gamma$, in conjunction with the execution intervals $EI_{i,k}, \forall \tau_{i,k} \in \tau_i$. However, the position of the execution intervals are tied to a schedule generated by a given scheduling policy. Without additional constraints to the task set, there is no flexibility with respect to changing the execution pattern of the tasks, because the execution pattern of the tasks is fixed according to the scheduling policy.

We consider that when a schedulable task set is available, we can change the order of execution of specific jobs. Therefore, we change the position and length of $I_i, \forall \tau_i \in \Gamma$, as well as the position and length of the execution intervals $EI_{i,k}, \forall \tau_{i,k} \in \tau_i$. One way to change the order of execution of specific jobs is by setting additional job-level dependencies (JLDs) to steer the scheduling algorithm, modify the schedule and obtain new values for $begin(I_i)$ and $end(I_i)$. We consider a dependency ($\prec$) between two jobs, $\tau_{i,k} \prec \tau_{j,l}$, if $\tau_{j,l}$ can only start its execution once $\tau_{i,k}$ has completed its execution. Thus, even if they are released at the same time, $\tau_{j,l}$ will only start its execution after $\tau_{i,k}$ has finished its execution.

Let us consider the schedule shown in Figure 2 and the following JLDs: $\{\tau_{2,0} \prec \tau_{1,0}, \tau_{1,0} \prec \tau_{3,0}, \tau_{2,2} \prec \tau_{3,3}\}$. By adding the JLDs to the schedule shown in Figure 2, we postpone the execution of $\{\tau_{1,0}, \tau_{3,0}, \tau_{3,3}\}$ and prepone the execution of $\{\tau_{2,0}, \tau_{2,2}\}$. As a consequence of adding precedence constraints between jobs of tasks $\tau_1$, $\tau_2$ and $\tau_3$, we rearrange the order in which jobs are executed. Therefore, we can compute new time points for $begin(I_i)$ and $end(I_i)$, which in turn result in new $I_i$ values. Figure 4 shows the new schedule of the tasks shown in Figure 2 when considering JLDs. Since the executions of $\tau_{1,0}$, $\tau_{3,0}$, and $\tau_{3,3}$ were postponed, and the executions of $\tau_{2,0}$ and $\tau_{2,2}$ were preponed, the values for $EI_{i,k}$, $\forall \tau_{i,k} \in \tau_i$, changed as shown in Figure 4. By using Equations 3 and 4, we compute new values $I_i, \forall \tau_i \in \Gamma$, as shown in Table 4.

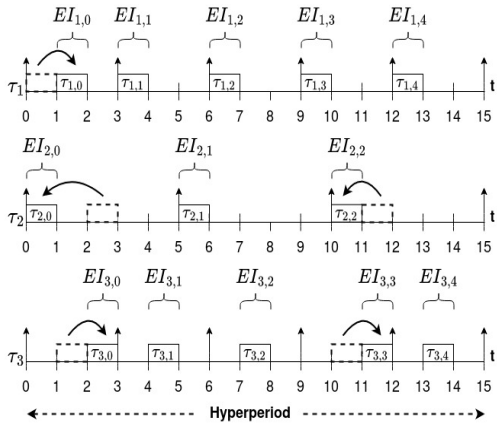

**Figure 4** $EI_{i,k}$ values obtained after adding JLDs

| Schedule-Aware Interval $I_i$ | $begin(I_i)$ | $end(I_i)$ |
| --- | --- | --- |
| $I_1$ | 0 | 2 |
| $I_2$ | 0 | 1 |
| $I_3$ | 1 | 3 |

**Table 4** $begin(I_i)$ and $end(I_i)$ values when the JLDs are considered

Figure 5 demonstrates how the new values for $begin(I_i)$ and $end(I_i)$ shown in Table 4 affect the E2E latencies and data propagation throughout the CEC $E$. Basically, Figure 5 shows that within the time interval of one $HP_E$, the LET model with schedule-aware intervals and JLDs produces 3 output values, in contrast to the models shown in Figures 1 and 3 that produce 1 and 2 output values respectively. Table 5 shows the E2E latencies for the first six outputs that propagated through the entire CEC $E$ based on inputs read by $\tau_1$ when $t \geq 0$. Note that the outputs produced at $t = \{3, 6\}$, which are based on values that did not propagate through the entire CEC are not shown in Table 5 because they do not represent complete E2E latencies.

By setting $L_i = I_i$, $\forall \tau_i \in \Gamma$, and adding JLDs between specific jobs, we can further reduce the worst-case data age of CEC $E$. Compared to the model proposed in Section 5.1 where no JLDs were considered, the worst-case data age reduced from 11 to 9, while the worst-case reaction latency increased from 8 to 9. However, the obtained value for the worst-case reaction latency is still lower than the worst-case reaction latency of 15 encountered with $L_i = T_i$. As long as the task set remains schedulable, additional JLDs can be added. As a consequence, it is possible to define different combinations for $I_i$ depending on how the execution order of the jobs is arranged according to the additional JLDs. There are multiple options to find the configuration of $I_i$ that best reduces the reaction latency and the data age. One possibility is to use an ILP solver or a search algorithm to find the most suitable set of $I_i, \forall \tau_i \in \Gamma$. Here, we implemented a search algorithm to find the set of additional JLDs that in conjunction with schedule-aware intervals best reduces the E2E latencies of multi-rate CECs. For the remainder of this paper, when considering the LET model with schedule-aware intervals and JLDs, we will refer to it as the Schedule-Aware LET model.

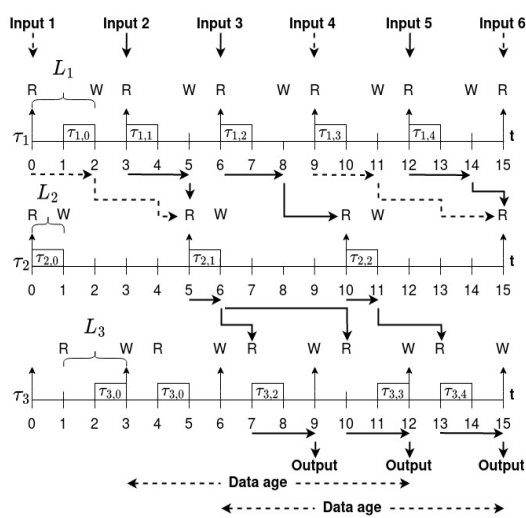

**Figure 5** Data propagation in $E$ with $L_i = I_i$ and additional JLDs

| Input at t: | Output at t: | E2E latency: |
|---|---|---|
| 3 | 9 | 6 |
| 3 | 12 | 9 |
| 6 | 15 | 9 |
| 12 | 18 | 6 |
| 12 | 21 | 9 |
| 18 | 24 | 6 |

**Table 5** E2E latencies of $E$ with $L_i = I_i$ and additional JLDs

We model our problem of reducing the E2E latencies of multi-rate CEC applying the Schedule-Aware LET model as a search tree and use a heuristic-based search algorithm to find a solution for it. We build our work on top a scheduling framework [19], which contains an improved version of the PIDA* algorithm. In our implementation, the root node of the search tree represents a feasible schedule of $\Gamma$ according to EDF. However, any scheduling policy that produces a feasible schedule of $\Gamma$ can be used. A child node represents a feasible schedule obtained after adding JLDs between two jobs. During each iteration of the search algorithm, it selects a job $\tau_{i,k}$ of $\tau_i$ and generates a new set of child nodes based on the JLDs added to $\tau_{i,k}$. According to our heuristic, we first investigate the child node that results in an $I_i$ that postpones $begin(I_i)$ and prepones $end(I_i)$. Schedule-aware intervals that have later time points for $begin(I_i)$ and earlier time points for $end(I_i)$ tend to result in shorter

E2E latencies. For each child node, we check for how many CECs the E2E latencies reduced after adding JLDs. The child node that improves the most number of CECs is selected as the next node to be investigated. We consider as a solution, the child node that reduces the E2E latencies of all the CECs present in the system. If no solution is found within a specific time interval defined by the user, the algorithm stops and returns among the searched nodes the set of additional JLDs that best reduce the E2E latencies.

In this section, we showed how the Schedule-Aware LET model can obtain less pessimistic E2E latencies of multi-rate CECs without compromising LET's deterministic characteristics.

## 6   Experimental Results

We evaluate our work based on the Real World Automotive Benchmarks presented by Kramer et al. in [12]. We compare our Schedule-Aware LET model with the approach presented in [4] and the LET model. The work in [4] will be referenced as the WCRT-LET model. We show the evaluation of our method for synthetic task sets in Section 6.2. Each simulation runs on a Intel(R) Core(TM) i7-9700 CPU @3.00GHz machine. We consider that the task sets run on a dual-core system.

### 6.1   Real World Automotive Benchmarks

We examined 1500 task sets based on the Real World Automotive Benchmarks [12]. Task periods are assigned to tasks as in [12] and the range of possible periods is: [1, 2, 5, 10, 20, 50, 100, 200, 1000]ms. Note that the sum of the probabilities for possible periods in [12] is 85%. The remaining 15% is for angle-asynchronous tasks. Since, angle-asynchronous tasks are not considered, all probability values are divided by 0.85. Inter-task communications follow the communication matrix shown in [12]. The total utilization of cores is $\approx 83\%$. On average, there are 40 CECs per task set. The number of tasks per task set is within the interval [80, 100], with an average of 92 tasks per task set. For this set of experiments, we let our algorithm run for 1 minute per task set. The obtained results are summarized on Figures 6 and 7. The box plots show a minimum 25% percentile, average, 75% percentile and the worst & best-case values. Results are normalized with respect to the worst-case data age in Figure 6 and worst-case reaction latency in Figure 7.

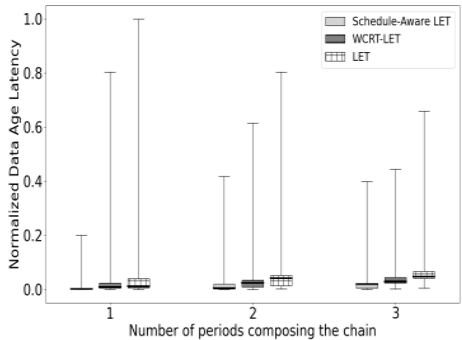 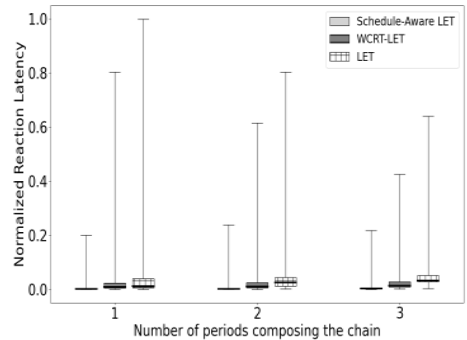

**Figure 6** Normalized data age       **Figure 7** Normalized reaction latency

Our Schedule-Aware LET model outperformed both WCRT-LET and LET models as shown in Figures 6 and 7. Our model managed to obtain worst-case data age values that are on average, ≈ 73% lower than the values obtained by LET as shown in Figure 8. On average, the WCRT-LET model managed to improve the worst-case data age values obtained by LET in ≈ 27%. As presented in [12], 70% of the CECs in the benchmarks are single-rate. Since our method shortens and reallocates the communication intervals of the LET model, input data can propagate throughout the CEC within one $HP_E$ rather than in multiple as with $L_i = T_i$. As a result, in some cases our method improved in over 90% the result obtained by the LET model. Figure 9 shows the improvements made by the Schedule-Aware LET model and the WCRT-LET model over LET with respect to worst-case reaction latency.

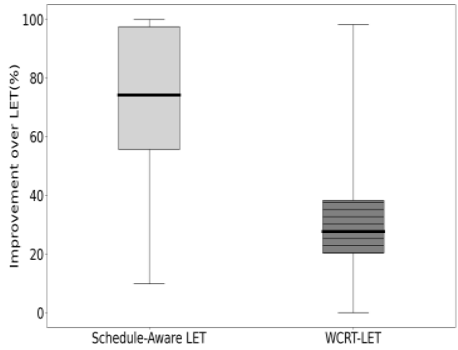
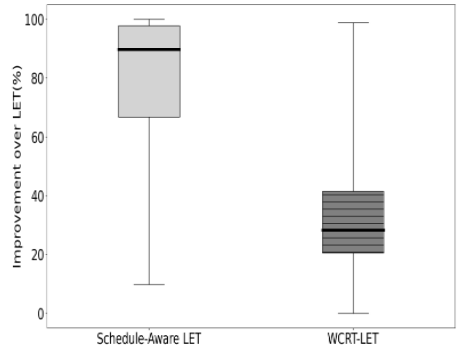

**Figure 8** Improvements over LET with respect to worst-case data age

**Figure 9** Improvements over LET with respect to worst-case reaction latency

## 6.2 Synthetically Generated Workloads

We examined 4000 randomly generated synthetic task sets based on the Real World Automotive Benchmarks [12]. However, we allowed tasks to have higher WCET values, increased the number of possible periods per CEC from 3 to 5 and drastically reduced the probability of single-rate CECs. The total utilization of cores is ≈ 76%. The number of tasks per task set is within the interval [30, 50], with an average of 40 tasks per task set. We let our algorithm run for 1 minute per task set. The obtained results are summarized in Figures 10 and 11.

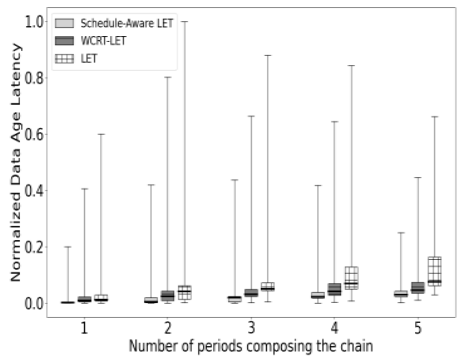
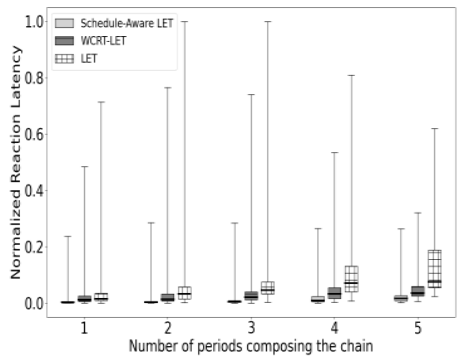

**Figure 10** Normalized data age

**Figure 11** Normalized reaction latency

As shown in Figures 10 and 11, our method outperformed WCRT-LET and LET in both metrics (data age and reaction latency). Figure 12 shows the improvements made by our Schedule-Aware LET model and the WCRT-LET model over the LET model. Our model obtained worst-case data age values that are on average, $\approx 65\%$ lower than the values obtained by LET. The WCRT-LET model improved worst-case data age values, on average, in $\approx 33\%$ compared to the LET model. Figure 13 shows the improvements made over the LET model with respect to the worst-case reaction latency.

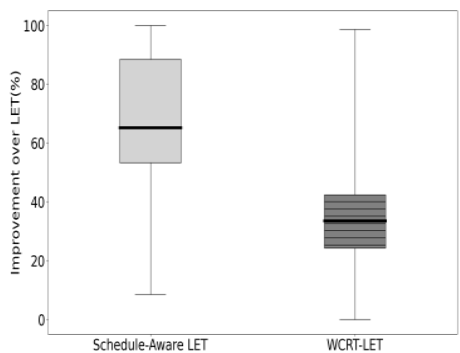

**Figure 12** Improvements over LET with respect to worst-case data age

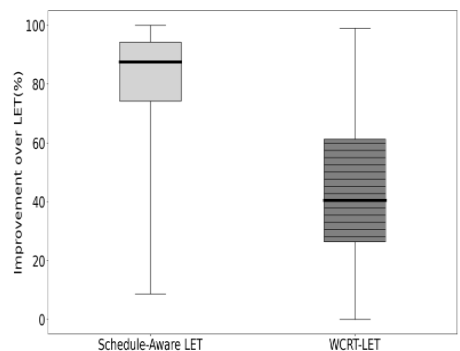

**Figure 13** Improvements over LET with respect to worst-case reaction latency

## 7  Conclusion and Future Work

In this paper, we proposed a method to obtain less pessimistic end-to-end latencies of multi-rate CECs applying the LET model, by considering more concrete knowledge of the scheduling algorithm in later design phases.

Starting from a feasible schedule, our method shortens and repositions communication intervals by making them schedule-aware, while maintaining the deterministic characteristics of LET. In addition, the method manipulates lengths and positions of the schedule-aware intervals by adding job-level dependencies between specific jobs to obtain less pessimistic end-to-end latencies.

Experiments showed that for task sets based on the Real World Automotive Benchmarks [12] or randomly generated, the Schedule-Aware LET model presented in Section 5 produced less pessimistic end-to-end latencies than the previously proposed models: WCRT-LET [4] and LET. For task sets based on the Real World Automotive Benchmarks [12], the Schedule-Aware LET model obtained worst-case data age and worst-case reaction latency values that are, on average, of $\approx 73\%$ and $\approx 89\%$ lower when compared to the LET model respectively. Moreover, if necessary for legacy reasons, the method presented in this paper does not have to be applied to the entire task set, but also to only a subset.

In future work, we plan to investigate the impact of mixing different execution models within the same CEC and how it can influence end-to-end latencies.